\input harvmac

\noblackbox


\def\IZ{\relax\ifmmode\mathchoice
{\hbox{\cmss Z\kern-.4em Z}}{\hbox{\cmss Z\kern-.4em Z}} {\lower.9pt\hbox{\cmsss Z\kern-.4em Z}}
{\lower1.2pt\hbox{\cmsss Z\kern-.4em Z}}\else{\cmss Z\kern-.4em Z}\fi}
\def\IB{\relax{\rm I\kern-.18em B}}
\def\IC{{\relax\hbox{\kern.3em{\cmss I}$\kern-.4em{\rm C}$}}}
\def\ID{\relax{\rm I\kern-.18em D}}
\def\IE{\relax{\rm I\kern-.18em E}}
\def\IF{\relax{\rm I\kern-.18em F}}
\def\IG{\relax\hbox{$\inbar\kern-.3em{\rm G}$}}
\def\IGa{\relax\hbox{${\rm I}\kern-.18em\Gamma$}}
\def\IH{\relax{\rm I\kern-.18em H}}
\def\II{\relax{\rm I\kern-.18em I}}
\def\IK{\relax{\rm I\kern-.18em K}}
\def\IP{\relax{\rm I\kern-.18em P}}

\font\cmss=cmss10 \font\cmsss=cmss10 at 7pt
\def\IR{\relax{\rm I\kern-.18em R}}

\def\frac#1#2{{#1 \over #2}}

\def\OL#1{ \kern1pt\overline{\kern-1pt#1
   \kern-1pt}\kern1pt }

\lref\KaloperSM{ N.~Kaloper, ``Bent domain walls as braneworlds,'' Phys.\ Rev.\ D {\bf 60}, 123506 (1999)
[arXiv:hep-th/9905210].
}

\lref\Padilla{ A.~Padilla, ``Brane world cosmology and holography,'' arXiv:hep-th/0210217;
J.~P.~Gregory and A.~Padilla, ``Exact braneworld cosmology induced from bulk black holes,'' Class.\ Quant.\
Grav.\  {\bf 19}, 4071 (2002) [arXiv:hep-th/0204218];
A.~Padilla, ``CFTs on non-critical braneworlds,'' Phys.\ Lett.\ B {\bf 528}, 274 (2002) [arXiv:hep-th/0111247].
 }

\lref\dks{ L.~Dyson, M.~Kleban and L.~Susskind,
JHEP {\bf 0210}, 011 (2002) [arXiv:hep-th/0208013].
}

\lref\GubserVJ{ S.~S.~Gubser, ``AdS/CFT and gravity,'' Phys.\ Rev.\ D {\bf 63}, 084017 (2001)
[arXiv:hep-th/9912001].
}

\lref\BalasubramanianZH{ V.~Balasubramanian, J.~de Boer and D.~Minic, ``Exploring De Sitter Space And
Holography,'' Class.\ Quant.\ Grav.\  {\bf 19}, 5655 (2002) [Annals Phys.\  {\bf 303}, 59 (2003)]
[arXiv:hep-th/0207245].
}

\lref\swconformal{
N.~Seiberg and E.~Witten, ``The D1/D5 system and singular CFT,'' JHEP {\bf 9904}, 017 (1999)
[arXiv:hep-th/9903224].
 }

\lref\DS{M.~Fabinger and E.~Silverstein,
``D-Sitter space: Causal structure, thermodynamics, and entropy,''
arXiv:hep-th/0304220.}

\lref\poincare{
L.~Dyson, J.~Lindesay and L.~Susskind,
``Is there really a de Sitter/CFT duality,''
JHEP {\bf 0208}, 045 (2002)
[arXiv:hep-th/0202163].}

\lref\lennyanth{L.~Susskind, ``The anthropic landscape of string theory,'' arXiv:hep-th/0302219.
}

\lref\KL{
D.~Kabat and G.~Lifschytz,
``Gauge theory origins of supergravity causal structure,''
JHEP {\bf 9905}, 005 (1999)
[arXiv:hep-th/9902073].}

\lref\adsdict{S.~S.~Gubser, I.~R.~Klebanov and A.~M.~Polyakov,
``Gauge theory correlators from non-critical string theory,''
Phys.\ Lett.\ B {\bf 428}, 105 (1998)
[arXiv:hep-th/9802109]. \
E.~Witten,
``Anti-de Sitter space and holography,''
Adv.\ Theor.\ Math.\ Phys.\  {\bf 2}, 253 (1998)
[arXiv:hep-th/9802150].}

\lref\andyjuan{
S.~Hawking, J.~M.~Maldacena and A.~Strominger, ``DeSitter entropy, quantum entanglement and AdS/CFT,'' JHEP {\bf
0105}, 001 (2001) [arXiv:hep-th/0002145].
}

\lref\otherandyjuan{J.~M.~Maldacena and A.~Strominger, ``Statistical entropy of de Sitter space,'' JHEP {\bf
9802}, 014 (1998) [arXiv:gr-qc/9801096].
}

\lref\counting{S.~Ashok and M.~R.~Douglas,
``Counting flux vacua,''
JHEP {\bf 0401}, 060 (2004)
[arXiv:hep-th/0307049]. \
}

\lref\martinec{
B.~C.~Da Cunha and E.~J.~Martinec,
``Closed string tachyon condensation and worldsheet inflation,''
Phys.\ Rev.\ D {\bf 68}, 063502 (2003)
[arXiv:hep-th/0303087].}

\lref\klebstrass{
I.~R.~Klebanov and M.~J.~Strassler,
``Supergravity and a confining gauge theory: Duality cascades and
$\chi$SB-resolution of naked singularities,''
JHEP {\bf 0008}, 052 (2000)
[arXiv:hep-th/0007191].}

\lref\otherdS{ E.~Witten, ``Quantum gravity in de Sitter space,'' arXiv:hep-th/0106109;
A.~Strominger, ``The dS/CFT correspondence,'' JHEP {\bf 0110}, 034 (2001) [arXiv:hep-th/0106113];
A.~Guijosa and D.~A.~Lowe, ``A new twist on dS/CFT,'' Phys.\ Rev.\ D {\bf 69}, 106008 (2004)
[arXiv:hep-th/0312282].
V.~Balasubramanian, J.~de Boer and D.~Minic, ``Exploring De Sitter Space And Holography,'' Class.\ Quant.\
Grav.\  {\bf 19}, 5655 (2002) [Annals Phys.\  {\bf 303}, 59 (2003)] [arXiv:hep-th/0207245].
}

\lref\tom{W. Fischler, ``Talk at 60th Birthday Celebration for G. West, June 2000"; T.~Banks, ``Cosmological
breaking of supersymmetry or little Lambda goes back to  the future. II,'' arXiv:hep-th/0007146;}

\lref\inprogress{work in progress}

\lref\junctions{E.~Silverstein,
``AdS and dS entropy from string junctions,''
arXiv:hep-th/0308175.
}

\lref\shahin{
M.~M.~Sheikh-Jabbari, ``Tiny graviton matrix theory: DLCQ of IIB plane-wave string theory, a conjecture,''
arXiv:hep-th/0406214.
}

\lref\kklt{ S.~Kachru, R.~Kallosh, A.~Linde and S.~P.~Trivedi, ``De Sitter vacua in string theory,''
arXiv:hep-th/0301240.
}

\lref\mss{ E.~Silverstein, ``(A)dS backgrounds from asymmetric orientifolds,'' arXiv:hep-th/0106209;
A.~Maloney, E.~Silverstein and A.~Strominger, ``De Sitter space in noncritical string theory,''
arXiv:hep-th/0205316.
}

\lref\tunneling{S.~Dimopoulos, S.~Kachru, N.~Kaloper, A.~E.~Lawrence and E.~Silverstein,
Phys.\ Rev.\ D {\bf 64}, 121702 (2001)
[arXiv:hep-th/0104239]; \
S.~Dimopoulos, S.~Kachru, N.~Kaloper, A.~E.~Lawrence and E.~Silverstein,
``Generating small numbers by tunneling in multi-throat  compactifications,''
arXiv:hep-th/0106128.}

\lref\Dccel{E.~Silverstein and D.~Tong,
``Scalar speed limits and cosmology: Acceleration from D-cceleration,''
arXiv:hep-th/0310221.}

\lref\DBIsky{M.~Alishahiha, E.~Silverstein and D.~Tong,
``DBI in the sky,''
arXiv:hep-th/0404084.}

\lref\karch{A.~Karch, ``Auto-localization in de-Sitter space,'' JHEP {\bf 0307}, 050 (2003)
[arXiv:hep-th/0305192]; talk at Texas A.\& M. String Cosmology Conference, March 2004.}

%

\lref\bms{R.~Bousso, A.~Maloney and A.~Strominger,
Phys.\ Rev.\ D {\bf 65}, 104039 (2002)
[arXiv:hep-th/0112218].}

\lref\multitrace{O.~Aharony, M.~Berkooz and E.~Silverstein,
``Multiple-trace operators and non-local string theories,''
JHEP {\bf 0108}, 006 (2001)
[arXiv:hep-th/0105309].\
E.~Witten, ``Multi-trace operators, boundary conditions, and AdS/CFT correspondence,'' arXiv:hep-th/0112258;
M.~Berkooz, A.~Sever and A.~Shomer, ``Double-trace deformations, boundary conditions and spacetime
JHEP {\bf 0205}, 034 (2002) [arXiv:hep-th/0112264].
}

\lref\nonabeliandbi{A.~A.~Tseytlin,
``On non-abelian generalisation of the Born-Infeld action in string  theory,''
Nucl.\ Phys.\ B {\bf 501}, 41 (1997)
[arXiv:hep-th/9701125].}

\lref\JuanDprobe{J.~M.~Maldacena,
``Branes probing black holes,''
Nucl.\ Phys.\ Proc.\ Suppl.\  {\bf 68}, 17 (1998)
[arXiv:hep-th/9709099].}

\lref\otherthermal{E.~Witten,
``Anti-de Sitter space, thermal phase transition, and confinement in  gauge
Adv.\ Theor.\ Math.\ Phys.\  {\bf 2}, 505 (1998)
[arXiv:hep-th/9803131].}

\lref\trapping{L.~Kofman, A.~Linde, X.~Liu, A.~Maloney, L.~McAllister and E.~Silverstein,
``Beauty is attractive: Moduli trapping at enhanced symmetry points,''
JHEP {\bf 0405}, 030 (2004)
[arXiv:hep-th/0403001].}

\lref\herman{H.~Verlinde,
``Holography and compactification,''
Nucl.\ Phys.\ B {\bf 580}, 264 (2000)
[arXiv:hep-th/9906182].}

\lref\gubpeet{
S.~S.~Gubser, I.~R.~Klebanov and A.~W.~Peet,
Phys.\ Rev.\ D {\bf 54}, 3915 (1996)
[arXiv:hep-th/9602135].
}

\lref\GKP{ S.~B.~Giddings, S.~Kachru and J.~Polchinski, ``Hierarchies from fluxes in string
compactifications,'' Phys.\ Rev.\ D {\bf 66}, 106006 (2002) [arXiv:hep-th/0105097].
}

\lref\RS{L.~Randall and R.~Sundrum,
Phys.\ Rev.\ Lett.\  {\bf 83}, 4690 (1999)
[arXiv:hep-th/9906064].}

\lref\ppnowave{A.~W.~Peet and J.~Polchinski,
``UV/IR relations in AdS dynamics,''
Phys.\ Rev.\ D {\bf 59}, 065011 (1999)
[arXiv:hep-th/9809022].
}

\lref\tseyank{A.~A.~Tseytlin and S.~Yankielowicz,
``Free energy of N = 4 super Yang-Mills in Higgs phase and non-extremal
D3-brane interactions,''
Nucl.\ Phys.\ B {\bf 541}, 145 (1999)
[arXiv:hep-th/9809032].
}

\lref\kirit{E.~Kiritsis and T.~R.~Taylor,
``Thermodynamics of D-brane probes,''
arXiv:hep-th/9906048.
}

\lref\kiritwo{
E.~Kiritsis,
``Supergravity, D-brane probes and thermal super Yang-Mills:  A comparison,''
JHEP {\bf 9910}, 010 (1999)
[arXiv:hep-th/9906206].
}

\lref\susswitt{ L.~Susskind and E.~Witten, ``The holographic bound in anti-de Sitter space,''
arXiv:hep-th/9805114.
}

\lref\buchel{A.~Buchel,
``Gauge / gravity correspondence in accelerating universe,''
Phys.\ Rev.\ D {\bf 65}, 125015 (2002)
[arXiv:hep-th/0203041]. \
A.~Buchel,
``Gauge / string correspondence in curved space,''
Phys.\ Rev.\ D {\bf 67}, 066004 (2003)
[arXiv:hep-th/0211141]. \
}

\lref\sussug{ L.~Susskind and J.~Uglum, ``Black hole entropy in canonical quantum gravity and superstring
theory,'' Phys.\ Rev.\ D {\bf 50}, 2700 (1994) [arXiv:hep-th/9401070].
}

\lref\lvm{ L.~Randall, V.~Sanz and M.~D.~Schwartz, ``Entropy-area relations in field theory,'' JHEP {\bf
0206}, 008 (2002) [arXiv:hep-th/0204038].
}

\lref\anothercitation{ J.~M.~Maldacena, ``The large N limit of superconformal field theories and supergravity,''
Adv.\ Theor.\ Math.\ Phys.\  {\bf 2}, 231 (1998) [Int.\ J.\ Theor.\ Phys.\  {\bf 38}, 1113 (1999)]
[arXiv:hep-th/9711200].
}

\lref\KovtunWP{
P.~Kovtun, D.~T.~Son and A.~O.~Starinets,
``Holography and hydrodynamics: Diffusion on stretched horizons,''
JHEP {\bf 0310}, 064 (2003)
[arXiv:hep-th/0309213].
}

\lref\PolicastroTN{
G.~Policastro, D.~T.~Son and A.~O.~Starinets,
``From AdS/CFT correspondence to hydrodynamics. II: Sound waves,''
JHEP {\bf 0212}, 054 (2002)
[arXiv:hep-th/0210220].
}

\lref\PolicastroSE{
G.~Policastro, D.~T.~Son and A.~O.~Starinets,
``From AdS/CFT correspondence to hydrodynamics,''
JHEP {\bf 0209}, 043 (2002)
[arXiv:hep-th/0205052].
}

\lref\HerzogPC{
C.~P.~Herzog,
``The hydrodynamics of M-theory,''
JHEP {\bf 0212}, 026 (2002)
[arXiv:hep-th/0210126].\
C.~P.~Herzog and D.~T.~Son,
``Schwinger-Keldysh propagators from AdS/CFT correspondence,''
JHEP {\bf 0303}, 046 (2003)
[arXiv:hep-th/0212072].\
C.~P.~Herzog,
Phys.\ Rev.\ D {\bf 68}, 024013 (2003)
[arXiv:hep-th/0302086].
}

\lref\gibbonshawking{
G.~W.~Gibbons and S.~W.~Hawking,
Phys.\ Rev.\ D {\bf 15}, 2738 (1977).}


\input epsf
\noblackbox
\newcount\figno
\figno=0
\def\fig#1#2#3{
\par\begingroup\parindent=0pt\leftskip=1cm\rightskip=1cm\parindent=0pt
\baselineskip=11pt \global\advance\figno by 1 \midinsert \epsfxsize=#3 \centerline{\epsfbox{#2}} \vskip 12pt
{\bf Fig.\ \the\figno: } #1\par
\endinsert\endgroup\par
}
\def\figlabel#1{\xdef#1{\the\figno}}

\Title{\vbox{\baselineskip12pt\hbox{hep-th/0407125} \hbox{SLAC-PUB-10540}\hbox{SU-ITP-04/29}
\hbox{IPM/P-2004/31}\hbox{MIT-CTP-3512} \hbox{UW/PT-04-07}}} {{\centerline{The dS/dS Correspondence}}}


\centerline{Mohsen Alishahiha$^{1,2}$, Andreas Karch$^3$, Eva Silverstein$^4$, and David Tong$^{4,5}$}
\bigskip

\centerline{\sl 1 Institute for Studies in Theoretical Physics and Mathematics, P.O. Box 19395-5531, Tehran, Iran}

\centerline{\sl 2 International Center for Theoretical Physics, 34100 Trieste, Italy}

\centerline{\sl 3 Department of Physics, University of Washington, Seattle, WA 98195}

\centerline{ \sl 4 SLAC and Department of Physics, Stanford University, Stanford, CA 94305/94309}

\centerline{ \sl 5 Center for Theoretical Physics, Massachusetts Institute of Technology, Cambridge, MA 02139}

\vskip .3in \centerline{\bf Abstract} {

We present a holographic duality for the de Sitter static patch which consolidates basic features of its
geometry and the behavior of gravity and brane probes, valid on timescales short compared to the decay or
Poincare recurrence times. Namely de Sitter spacetime $dS_d(R)$ in $d$ dimensions with curvature radius $R$ is
holographically dual to two conformal field theories on $dS_{d-1}(R)$, cut off at an energy scale $1/R$ where
they couple to each other and to $d-1$ dimensional gravity. As part of our analysis, we study brane probes in de
Sitter and thermal Anti de Sitter spaces, and interpret the terms in the corresponding DBI action via strongly
coupled thermal field theory. This provides a dual field theoretic interpretation of the fact that probes take
forever to reach a horizon in general relativity.


} \vskip .1in

\smallskip
\Date{July 2004}

\listtoc
\writetoc


\newsec{Introduction}

In this paper we propose a holographic, dual description of the de Sitter static patch, along the lines of
\refs{\DS,\karch,\junctions}. This relationship is analogous to that arising in warped compactifications or
Randall-Sundrum geometries \refs{\RS,\herman,\GKP,\GubserVJ}\ with multiple throats \tunneling. We motivate and
study the duality through the use of gravity and brane probes. Before delving into the details, let us start by
explaining the main points.

The static patch in $d$ dimensional de Sitter space of radius $R$ can be foliated by $dS_{d-1}$ slices:
\eqn\metslice{ds^2=\sin^2\left({w\over R}\right)ds^2_{dS_{d-1}}+dw^2}
The resulting metric \metslice\ has a warp factor which is maximal with finite value at a central slice $w =\pi
R/2$, dropping monotonically on each side until it reaches zero at the horizon $w=0,\pi R$ (see figure 1). The
region near the horizon, which corresponds to low energies in the static coordinates, is isomorphic to that of
$d$ dimensional $AdS$ space foliated by $dS_{d-1}$ slices (for which the warp factor is $\sinh^2(w/R)$ rather
than $\sin^2(w/R)$) and hence constitutes a CFT on $dS_{d-1}$ at low energies. Correspondingly, D-brane probes
of this region exhibit the same rich dynamics as a strongly coupled CFT on its approximate Coulomb branch.
Meanwhile, probes constructed from bulk gravitons range from energy 0 up to energy $1/R$ at the central slice,
and upon dimensional reduction their spectrum exhibits the mass gap expected of $d-1$ dimensional conformal
field theory on de Sitter space. Dimensionally reducing to the $d-1$ dimensional effective field theory also
yields a finite $d-1$ dimensional Planck mass, so the lower dimensional theory itself includes dynamical
gravity. As we shall see, the value of this lower dimensional Planck mass is consistent with that generated by a
renormalization of Newton's constant from $S$ degrees of freedom cut off at the scale $1/R$ (where $S$ is the
horizon area in Planck units).  Altogether, the geometry and energy scales are as summarized in figure 1.
%
%
\global\advance\figno by1
\ifig\spatialfig{A spatial slice of the static patch of $dS_d$, with the $dS_{d-1}$ slices and the behavior of
the redshift factor $g_{00}$ shown. The bulk of the $d$ dimensional spacetime is described by conformal field
theories on $dS_{d-1}$ at energy scales $E$ ranging from $0$ to of order $1/R$, where they are cut off and
coupled to each other and to $d-1$ dimensional gravity, which is localized at the central slice where
$g_{00}=1$.  We have indicated $dS_{d-1}$ slices near the IR end on each side, on which we can consider brane
probes realizing the approximate Coulomb branch of the low energy CFTs.} {\epsfxsize2.5in\epsfbox{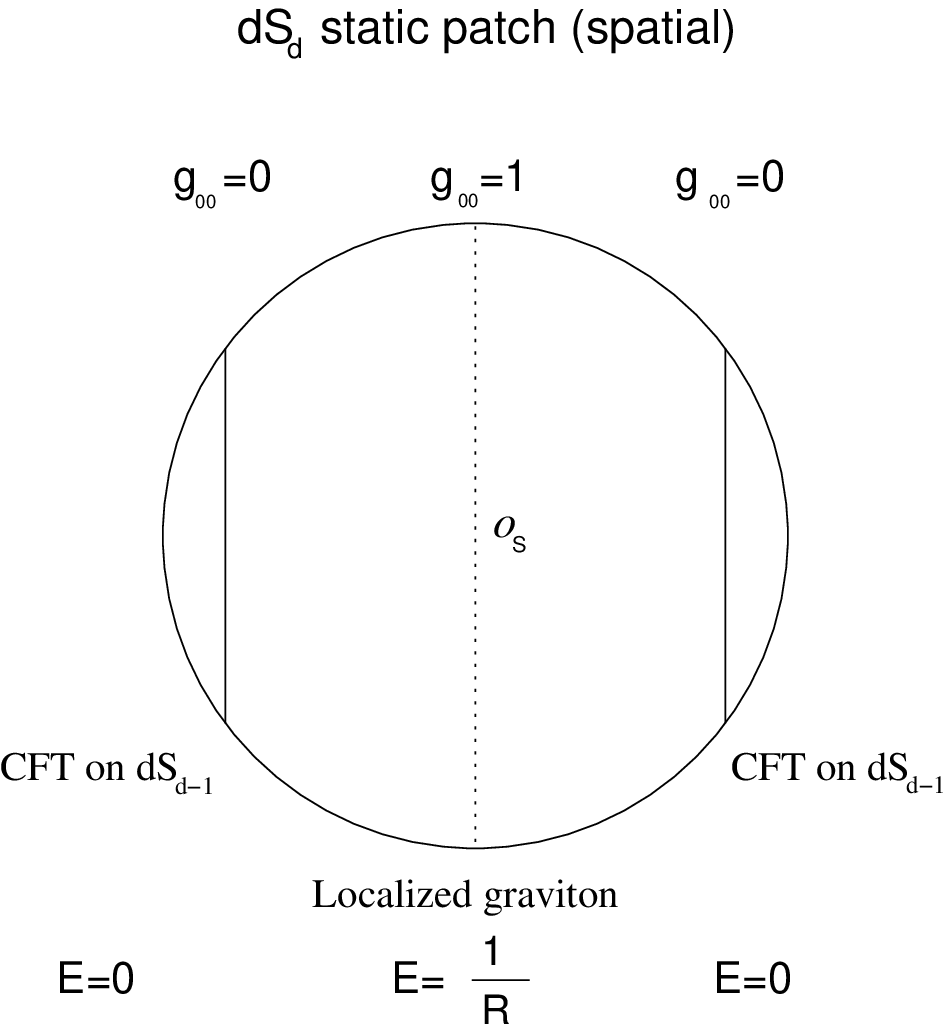}}

This leads to the following statement of de Sitter holography.  The $dS_d$ static patch is dual to two conformal
field theories on $dS_{d-1}$ (hence thermal with temperature $T=1/R$), cut off at an energy scale $1/R$ and
coupled to each other as well as to $(d-1)$-dimensional gravity.

In principle it should be possible to iterate this procedure to successively higher codimension \karch\ using
the explicit field theoretic degrees of freedom obtained in \refs{\DS,\junctions}\ in string theoretic models
such as \refs{\mss,\kklt}, culminating in a quantum mechanical description independent of any gravitational
sector; we leave this program for future work \inprogress.  The existence of the cut-off on energies -- even if
we descend all the way to quantum mechanics -- provides us with only a finite window of accessible states of the
dual theory, a requirement which has been stressed particularly in \tom, and noted in the context of the static
patch in \dks\ where the importance of a self-contained holographic description of a single observer's
accessible region was emphasized. Also, although here we will focus on timescales short compared to the decay
and Poincare recurrence times \refs{\kklt,\mss,\poincare,\dks}, we will see that the non-perturbative decay out
of the de Sitter phase is mirrored in the dual via vacuum bubble nucleation.

As it stands, restricting ourselves to times short compared to the decay time, and codimension one, this type of
duality is analogous to the holographically dual description of the Randall-Sundrum or warped compactification
geometries or, more precisely, the multi-throated versions studied in
\refs{\tunneling, \GKP}. That is, the
$d$-dimensional gravity theory is dual to $d-1$ dimensional cut off quantum field theory coupled to gravity.
Both sides of the correspondence involve gravity (in different dimensions); nonetheless the holographic relation
is useful because the bulk of the entropy is carried by the field theoretic degrees of freedom.  This analogous
RS case is reviewed in figure 2.

\ifig\RSfig{The two-throated Randall-Sundrum or warped compactification depicted here has a similar holographic
duality: the dual is given by a pair of field theories cut off in the UV and coupled to each other and to $d-1$
dimensional gravity.}{\epsfxsize5.0in\epsfbox{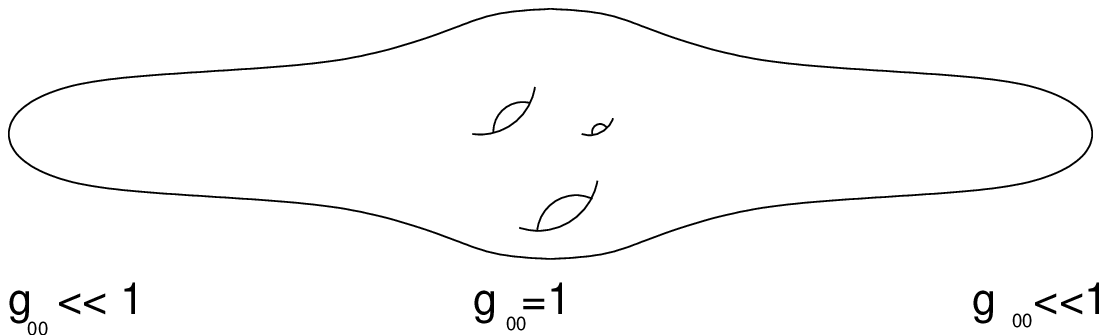}}

In the bulk of this paper, we will analyze the D-brane and gravity probes of these geometries, and the related
case of $dS$-sliced $AdS$ (which is isomorphic to our case at low energies). The gravity probe analysis yields a
localized graviton (as well as a finite set of other localized modes) and a continuum above a gap of ``glueball"
modes as in \karch, as well as results on the hydrodynamical behavior of the field theory in its low energy
regime. The brane probe analysis contains information about the microphysics of this dual field theory regime
out on its Coulomb branch as in \refs{\DS,\junctions,\Dccel,\DBIsky}.

In addition to fleshing out the basic structure of the correspondence summarized above, this leads to
interesting lessons about strong coupling phenomena.  For instance, in general relativity, from the point of
view of a static, outside observer, it takes forever for a probe to reach a horizon.  In quantum field theory, a
rolling scalar field experiences large back reaction on its motion as it approaches a point with new light
states \refs{\trapping,\Dccel,\DBIsky}.  These two phenomena are holographically dual, as studied recently for
zero temperature horizons in \refs{\KL,\Dccel,\DBIsky}.
Our analysis in this paper demonstrates how this dual
description arises for probes of more general horizons with nonzero area.

\newsec{de Sitter slicing of $(A)dS_d$ and de Sitter holography}

In this section we will explain the structure of the geometry and gravity probe behavior entering into the
holographic relation for de Sitter space summarized above.  In the following two sections we will study the
D-brane probe dynamics which encodes basic features of the low energy field theory regime of the dual.  These
analyses will provide some of the details of the physical phenomena arising in this system from the dual points
of view; it is worth emphasizing that the isomorphism of the low energy region described above is the basic
reason for the existence of the duality (as in the original arguments for $AdS/CFT$ \anothercitation).

We will study both $dS_d$ and $AdS_d$ in parallel, each foliated by $dS_{d-1}$ slices (see
\refs{\KaloperSM,\Padilla}\ for some early work on such foliations in the context of the Randall-Sundrum
scenario). In the case of $AdS_d$, the gauge-gravity correspondence in this foliation has been discussed in
\refs{\andyjuan,\buchel}, and results in a $d-1$ conformal field theory on $dS_{d-1}$ which therefore
necessarily inherits some thermal properties. As we shall see, the $dS_d$ case is isomorphic to the $AdS_d$ case
at leading order at low energy, enabling us to identify the near-horizon region of the bulk of the $dS_d$ static
patch also as $d-1$ dimensional conformal field theory on $dS_{d-1}$ at energy scales below a cutoff scale
determined by the geometry.

The metric for a $dS_{d-1}$ slicing of $(A)dS_d$ is
\eqn\otherdSmet{ds^2=e^{2A(w)}\hat g_{ij}dy^idy^j+dw^2}
where $\hat g_{ij}$ is a metric for $dS_{d-1}$, and the warp factor is given by $e^{2A(w)}=\sin^{2}(w/R)$ in the
case of $dS_d$, while for $AdS_d$, we have $e^{2A(w)}=\sinh^2(w/R)$. In the $dS_d$ case, we will work within a
single causal patch. In both cases, we may change coordinates to $\exp(2A(w))=\rho^2/R^2$, to write the metric
in the form
\eqn\dSslices{ds^2={\rho^2\over R^2}\hat g_{ij}dy^idy^j + {R^2\over {R^2\pm \rho^2}}{d\rho^2} }
where the $\pm$ is $+$ for $AdS_d$ and $-$ for $dS_d$. In the $AdS_d$ case, we are by now used to replacing the
radial direction $\rho$ with the energy scale of a dual $d-1$ conformal field theory. Here we wish to do the
same for $dS_d$. Let us examine the consequences in the IR and UV.

The IR of the dual theory corresponds to the near horizon regime $\rho=0$ where the two metrics \dSslices\
coincide at leading order. As we shall describe, this implies that the resulting DBI actions describing the
Coulomb branch brane probe dynamics in this region also agree. From the $d-1$ perspective, the energy of
(super)-gravity modes localized at some position $\rho$ is given by the warped down Kaluza-Klein scale in the
geometry
\eqn\GRdSlice{E\sim {1\over R}\sqrt{g_{00}}\sim {\rho\over R^2}}
where the curvature scale $1/R$ of the $dS_{d-1}$ also provides the temperature scale of the $d-1$-dimensional
field theory (taking the Euclidean vacuum for the CFT on $dS_{d-1}$) \gibbonshawking. So we see that at leading
order in a low energy expansion, that is neglecting the corrections of order $\rho^2/R^2=E^2/T^2$, the $dS_d$
case can also be identified with conformal field theory on $dS_{d-1}$ in the same manner as $AdS_d$.

In both cases, the number of degrees of freedom of the field theory is of order $S=R^{d-2}M_d^{d-2}$ where $M_d$
is the $d$-dimensional Planck mass, and the counting is the Gibbons-Hawking entropy for $dS_d$ \gibbonshawking,
and the Susskind-Witten entropy for $AdS_d$ \susswitt. Below we will study in more detail some of the structure
of this low energy theory using gravity and brane probes.  For the moment we simply emphasize that the results
of this analysis are identical at low energies for the $AdS_d$ and $dS_d$ cases.


At higher energies, the two spaces \dSslices\ become distinct.  The $AdS_d$ case asymptotes to a boundary at
infinite $\rho$.  The $dS_d$ case of the metric \dSslices\ exhibits a coordinate singularity at $\rho=R$;
returning to the form \otherdSmet\ reveals this as the central $dS_{d-1}$ slice of the $dS_d$ static patch, at
which $g_{00}$ is maximal and the GR probe energy scale reaches $1/R$.  On the other side of this slice there is
another bulk region, identical to the first and also identifiable with low energy CFT on $dS_{d-1}$ (see figure
1).  The holographic description of the $dS_d$ static patch therefore includes two CFTs, cut-off at energy scale
$1/R$ where they couple to each other. Note that we are discussing a single observer's static patch and the two
CFTs are in causal contact with each other, rather than in an entangled state as occurs in some other
descriptions of thermal de Sitter systems (see for example \andyjuan).

In addition, the $dS_d$ geometry differs from the $AdS_d$ geometry by the presence of a normalizable zero mode
for the $d$-dimensional graviton, resulting in dynamical gravity (i.e. with a finite Planck mass) in the $d-1$
dimensional reduction \karch.  The value of the Newton constant $G_N$  determined in the fully gravitational
description via dimensional reduction from $d$ to $d-1$ dimensions is
\eqn\rennewtII{{1\over G_{N,d-1}}\sim M_{d-1}^{d-3}\sim RM_d^{d-2}\sim(RM_d)^{d-2}\left({1\over R}\right)^{d-3}\sim
S\left({1\over R}\right)^{d-3} }
Naively this $d-1$ dimensional Planck scale is much higher than the UV cut-off of the CFTs, which might give us
pause. However, as happens in Randall-Sundrum geometries, the scales in the problem are consistent with the
interpretation of the contribution to the $d-1$ Newton constant from the warped region  coming from a
renormalization in the dual field theory. The importance of this effect was first emphasised in \sussug\ in the
context of black holes (see also \refs{\andyjuan,\lvm}\ for applications in RS scenarios).\foot{We thank R.
Sundrum for a useful discussion of related issues.} With $S$ propagating low-energy field modes with cut-off
$M_{UV}=1/R$, the renormalised Newton constant in the $d-1$ dimensional effective field theory given by
\eqn\rennewtI{{1\over G_{N,d-1}}\sim S M_{UV}^{d-3}\sim S \left({1\over R}\right)^{d-3}}
which coincides with that derived from dimensional reduction along the warped throat. This general structure
obtained in the $dS_d$ case would also arise in the $AdS_d$ case if we considered two $AdS_d$ throats cut off
and joined at their UV ends with a Randall-Sundrum ``Planck brane" or equivalently a warped compactification
\refs{\herman,\GKP,\tunneling,\kklt,\andyjuan}.

Altogether, this leads to a holographic description of the $d$-dimensional de Sitter static patch, valid on
timescales shorter than the recurrence and/or decay time.\foot{As we will see in \S4.3, in situations where de
Sitter decays there is a dual description of this as well.} It is conformal field theory on $d-1$-dimensional de
Sitter space, thus endowed with a temperature $1/R$, cut off at the scale $1/R$, and coupled to gravity and
another conformal field theory. In the following, we elucidate some salient features of the $d-1$ dimensional
holographic dual.

\subsec{Symmetries}

The symmetry of the $dS_d$ static patch includes $SO(d-1)$, time translations, and reflections. The
corresponding symmetry transformations implement a UV/IR relation in our system.  For example, taking a
localized probe at some $w$ and $y$ position on say the left side of figure 1, and rotating it to a higher $w$
position using the $SO(d-1)$ symmetry also changes its position $y$ in $dS_{d-1}$ so that it is closer to the
$dS_{d-1}$ horizon.

\subsec{Kaluza-Klein Modes}

Note that in this correspondence, the energy scales $E \ll T$ are crucial to the utility of the holographic
duality; these are the scales in which the pure field theory description applies.  Let us clarify the energy
scales $E$ in the $dS_{d-1}$ theory, including those coming from excitations of the massless $dS_d$ fields. As
discussed above, the redshifting of static energy scales by the $\sqrt{g_{00}}$ factor in the metric leads to a
continuum of energies $E$ starting at zero. Modes localized near the horizon fall in toward the horizon, a
process which represents the equilibration of these excitations in the thermal field theory. On the other hand,
the slicing of $dS_d$ by $dS_{d-1}$ slices leads to a spectrum of masses in the $d-1$ dimensional theory
including a continuum above a gap (as well as the zero mode and some discrete modes whose role we will come to
presently) \karch.

The gravity modes on these backgrounds can be studied by reducing the wave equation to an analogue Schr\"odinger
equation as in \refs{\RS,\karch}.  This analogue Schr\"odinger problem is
\eqn\Schrod{-\psi'' + \frac{1}{R^2} \left ( \frac{(d-2)^2}{4} - \frac{(d-2)}{4} \frac{d}{\cosh^2(z/R)}
\right )\psi = m^2\psi.}
Here $\psi$ is the $d$ dimensional transverse traceless graviton or scalar field rescaled by a factor which
reduces to $e^{-(d-2)^2z/4R^2}$ at large $z$ where $z$ is the radial coordinate for the metric in the form
$ds^2_{dS_d}=e^{2A(z)}(ds^2_{dS_{d-1}}+dz^2)$, and $^\prime$ represents a derivative with respect to this
conformal radial coordinate $z$. This rescaling is such that the norm on $\psi$ is $\int |\psi|^2$ as in the
analogue quantum mechanics problem defined by \Schrod.  Very near the horizon $\rho\to 0$, $z$ approaches
infinity and the potential term in \Schrod\ approaches a constant $V\to {(d-2)^2/{4R^2}}$ This yields a
continuum of modes above a mass gap
\eqn\massgap{m^2 > {(d-2)^2\over{4R^2}}}
Note that this mass gap does not imply an energy gap in the $d-1$-dimensional theory (whose energies were given
above in \GRdSlice). If we work in static coordinates in the $d-1$ dimensional theory, we can place the massive
particles near the horizon in the $d-1$ dimensional de Sitter space, leading to arbitrarily small-energy
excitations.  The $\sqrt{g_{00}}$ factor in the $(A)dS_d$ goes to zero as $\rho\to 0$ regardless of the
coordinates we use to describe the $dS_{d-1}$, so the presence of small energy scales in the $d-1$ dimensional
effective field theory should be more general. Indeed, in global coordinates on $dS_{d-1}$, the mode solutions
of \bms\ have a real part in their frequency $\mu$ (in the notation of \bms) precisely when the equation
\massgap\ is satisfied. Finally, let us note that the mass gap here \massgap\ is precisely the same as that
obtained for a $dS_{d-1}$ slicing of $AdS_d$, i.e. it is the mass gap arising in conformal field theory on de
Sitter space.

In addition to the continuum, the $1/\cosh^2$ potential, also known as the P\"oschl-Teller potential,
exhibits bound states. Most notably there is a
normalizable zero mode, like in the two-throated RS scenario. This is crucial for our interpretation of dS
duality as two cutoff conformal field theories, coupled to each other and to lower dimensional gravity at the
cutoff scale $1/R$.

In general, the zero mode is not the only bound state. The P\"oschl-Teller
potential, can be solved analytically and the bound states are labelled by $l=0,1,\ldots$ with $l <
\frac{d-2}{2}$ and have masses
\eqn\extrastates{ m_l^2 = - \left ( \frac{d-2}{2R} - {l\over R} \right )^2+\frac{(d-2)^2}{4R^2}. }
%
They are associated to the UV brane, in the same way that warped compactifications can have light modes
localized in the region outside the warped throats.  As such they encode details about the way the two throats
are cut-off and coupled to each other. It would be interesting to decode their effects in detail. Note that the
first extra mode shows up in $d=5$, and as a bound state it translates into a mode of negative $\mu^2$.

\subsec{Hydrodynamics}

Another aspect accessible to gravity probes is the hydrodynamics of the thermal field theory. Hydrodynamics is
the effective description for low energy, long wavelength fluctuations around equilibrium. Since in the thermal
field theory dual to dS we are forced into the low energy regime, long wavelength fluctuations are described by
hydrodynamics.

Many hydrodynamic processes have been uncovered from the dual gravity in
\refs{\PolicastroTN,\PolicastroSE,\KovtunWP,\HerzogPC}\ using purely outgoing boundary conditions on the horizon
to reflect the dissipative nature of the diffusion process. One then either looks at correlation functions of
the hydrodynamic variables, that is the conserved currents, or at the spectrum of quasinormal modes. From this
one can extract the speed of sound, bulk and shear viscosity and the diffusion constants of any currents. In a
locally flat region, the diffusion constant $D$ for a conserved current associated to an abelian gauge field is
given by a very simple universal formula obtained in \KovtunWP
\eqn\diffusion{D=\frac{\sqrt{-g(w_0)}}{g_{xx}(w_0) \sqrt{-g_{00}(w_0) g_{ww}(w_0)}} \; \; \int_{w_0}^{\pi/2}
\frac{- g_{00}(w) g_{ww}(w)}{\sqrt{-g}}}
where $w_0$ is introduced as the position of the stretched horizon, which in the end we take to coincide with
the dS horizon $w=0$ in the dS slicing \metslice.
To extract the diffusion coefficient, we are interested only in the $w$ dependence of the metric factors (as
opposed to contributions due to the fact that the diffusion process is itself occurring on a curved $dS_{d-1}$
background). Plugging in the metric \metslice\ for $d=5$ we get $D=R$, which is consistent with the thermal
interpretation. (Note that had we done the same computation in a radial slicing of the $dS_5$ static patch, as
opposed to the $dS_4$ slicing we are considering here, we would have obtained an infinite answer). In $d>5$ with
the $dS_{d-1}$ slicing one similarly obtains a finite answer, while in $d=4$ the diffusion constant is log
divergent, reflecting the behavior of hydrodynamics on the plane.

\subsec{Physics at the cutoff and deformations}

The matching of the full theory to the CFT at the scale $1/R$ should be determined -- again as in the warped
compactification cases -- via the behavior of gravity modes in the static patch near the central UV slice. It
will be interesting to work the details of this out.  Since we have not done so, our statement of holographic
duality is not complete  in the same way that the oft-repeated statement ``Randall-Sundrum is just CFT coupled
to gravity" is not complete.  However, as in that case \GubserVJ\ we believe the statement is useful and
interesting as it stands.

Relatedly, as with the $AdS/CFT$ correspondence and its application to warped compactifications, there are many
variants and deformations of this basic setup, some already familiar. In particular, quantum field theories on
de Sitter space cut off and coupled to gravity also arise as the holographic duals to more complicated flux
compactifications.  Consider for example the KKLT models of $dS_4$ space \kklt, which involve a warped throat
emanating from a Calabi-Yau compactification of type IIB string theory.  This warped throat is dual to a field
theory (in the specific examples considered in \kklt\ this was a Klebanov-Strassler cascading gauge theory
\klebstrass) living on $dS_4$, at energy scales ranging up to a finite cutoff scale determined by matching this
throat to the Calabi-Yau whose finite volume ensures a nontrivial coupling to four dimensional gravity.  So this
set of field theories formulated on $dS_4$, cutoff and coupled to gravity in the specific way determined by the
Calabi Yau, is not dual to $dS_5$ but to the warped flux model (a duality also studied in
\refs{\andyjuan,\buchel}).

The difference between a pair of cutoff field theories dual to $dS$ and those dual to a warped compactification
model lies in part in the interactions at the cutoff scale determined by the geometry in the UV region. In some
sense, the $dS$ case is the simplest and most symmetric, and the warped compactification models are more
complicated and more generic.

An interesting intermediate situation which we encountered in our investigation is the following, illustrating
the kind of generic deformation of our duality that we expect. Consider a warped de Sitter compactification such
as \kklt, but with the added specification that the fields in the field theory dual to the warped throat are
thermally excited at a temperature $T$ distinct from the Hubble temperature scale of the de Sitter.  This should
lead to a thermal horizon in the warped throat of the compactification, whose radial position $r_0$ corresponds
to the energy scale of the thermal bath of excited field theory modes (as we will review in the case of pure
thermal field theory in \S3). Since the accelerated expansion of the de Sitter compactification dilutes this
thermal bath, the corresponding energy density should decrease with time, and hence the radial position $r_0$ at
which the geometry is modified due to the thermal bath should recede.  As a result, such a system includes an
expanding $5th$ dimension, much like the $dS_5$ case of interest in this paper.

\subsec{Brane Probes}

Finally, we come to D-brane probes, a subject which shall occupy us for the next two sections. The DBI action
for the brane contains higher derivative terms which encode the effect of microscopic interactions in the dual
strongly coupled field theory, as seen in the context of AdS/CFT in e.g. \refs{\JuanDprobe,\anothercitation}.
Our goal will in part be to disentangle the origin of these terms. As we shall see, in general they arise from a
combination of radiative corrections arising from the coupling of the probe to the rest of the degrees of
freedom, together with an evaluation of these degrees of freedom in a thermal ensemble. An analysis of these
terms can therefore yield microscopic information about the entropy and couplings of the full system. Before
considering the probes on $dS_{d-1}$ slices, we first consider the somewhat simpler case of thermal CFT on
Minkowski space, whose lessons we will subsequently apply to de Sitter in Section 4.

\newsec{AdS Black Brane Probes}

In this section we shall study a probe D3-brane in Schwarzchild $AdS_5$ geometry. The main goal is to explain
how the expansion of the DBI action may be interpreted in terms of thermal field theory, as a prelude to the
case of brane probes in the de Sitter slicing of interest for our duality. We shall show that the various terms
can be understood as higher dimension operators evaluated in a thermal bath. Some aspects of this interpretation
were discussed in \refs{\tseyank,\kirit,\kiritwo,\otherthermal}\ as well as in \JuanDprobe, and we shall extend
their results.

The static patch of the $AdS_5$-Schwarzchild space can be obtained by taking the near horizon limit of $N$ black
D3-branes in type IIB supergravity, resulting in the metric \eqn\adsfive{ ds^2={1\over
f(r)^{1/2}}(-h(r)dt^2+d\vec{x}\cdot d\vec{x})+{f(r)^{1/2}\over h(r)}dr^2 } where $f(r)=R^4/r^4$ and the function
$h(r)=1-r_0^4/r^4$ parameterizes the deviation from extremality. The horizon lies at $r=r_0$. The background is
accompanied by a constant dilaton $e^\phi=g_s$ and a non-vanishing five-form RR flux, with potential
$C_4=f(r)^{-1}dt\wedge dx_1\wedge dx_2\wedge dx_3$.

This supergravity background is dual to $SU(N)$ ${\cal N}=4$ super Yang-Mills at finite temperature. The gauge
coupling constant is given by $g^2_{YM}\sim g_s$, while the 't Hooft coupling $\lambda=g^2_{YM}N\sim (R/l_s)^4$
is necessarily large for the supergravity description to be valid. The temperature $T$ of the dual field theory
coincides with the Hawking temperature of the black hole: $T\sim r_0/l^2_s\sqrt{\lambda}$. As is well known,
analysis of this supergravity background can be used to extract information about the thermodynamics of the
strongly coupled gauge theory. For example, the theory can be shown to have free energy given by ${\cal F}\sim
N^2 T^4$ (where a more careful analysis, including numerical coefficients, reveals the famous relative factor of
$3/4$ between the free-energy at weak and strong coupling \gubpeet).

The metric exhibits the UV/IR relationship which associates radial locations with energy scales in the field
theory. In fact, as explained by Peet and Polchinski \ppnowave, there are at least two different energy scales
associated to a given radial position $r$. The first arises from supergravity modes localised at $r$ and is the
relevant scale for computing the entropy \susswitt. The energy of these excitations, which we shall refer to as
GR probes, is given by the Kaluza-Klein energy scale $1/R$, warped down by the red-shift factor $\sqrt{g_{00}}$,
\eqn\kkscale{ E\sim {1\over R}\sqrt{g_{00}}\sim {r\over R^2}\sqrt{h} }
The second energy scale is associated to stretched strings which, following \trapping, we refer to as
``$\chi$-modes''. These appear on the (thermally lifted) Coulomb branch of the field theory and are described on
the gravity side as a string stretched between the horizon and a probe D3-brane localized at position $r$. These
have mass
\eqn\mchi{ m_\chi\sim\
{1\over l_s^2}\int dr \sqrt{g_{00}}\sqrt{g_{rr}} \sim {r\over l_s^2} }
which, at high energies, exceeds the KK scale: $m_\chi\approx E\sqrt{\lambda}$. The presence of the probe
D3-brane acts as a domain wall in the $5d$ gravity background, jumping the $F_5$ RR-flux by a single unit. It
corresponds to a scalar field VEV implementing gauge symmetry breaking $U(N)\rightarrow U(1)\times U(N-1)$.

Let us delineate the $U(N)$ degrees of freedom of the field theory as follows: there is the single $U(1)$ eigenvalue
$\phi$ which, at zero temperature, is related to the brane position by $r=\phi l_s^2$. The bulk of the system is
made up of the $U(N-1)$ degrees of freedom which we will collectively refer to as $\eta$. These are coupled to
$\phi$ through the off-diagonal $\chi$ modes. The effective action for the probe degree of freedom $\phi$ has a
rich structure encapsulated in the DBI action. We set the field strength on the probe brane to zero and
concentrate only on the radial fluctuations, resulting in the action
\eqn\dbiads{ {\cal L}_{DBI}=-{1\over {g_sl_s^4}}{1\over f(r)}\left(h(r)^{1/2}\sqrt{1-{f(r)\over h(r)^2}
\dot{r}^2+{f(r)\over h(r)}(\nabla r)^2}-1\right) }
Our goal here is to interpret this as arising from interactions with the $\eta$ and $\chi$ degrees of freedom.
We will start with a discussion of the general structure of the couplings implied by the probe DBI, and then
turn to a discussion of its behavior in the limits of $T/E \ll 1$ and $T/E \gg 1$.

\subsec{General Structure}

In the dual field theory, the form of the DBI action \dbiads\ arises from a two-step process: firstly one
integrates out the $\chi$ modes, resulting in an effective action for the $\phi$ and $\eta$ fields. Secondly,
all $\eta$ operators are evaluated in the thermal background. At zero temperature, the DBI action for $\phi$ is
obtained by integrating out the $\chi$ and $\eta$ modes, and a non-renormalization theorem ensures that the form
of the DBI action can be successfully compared with the effective action for $\phi$ even at weak coupling
\anothercitation\ (although at weak coupling particle production effects dominate over the DBI corrections for a
rolling scalar field \trapping). At finite temperature it is necessary to include the $\eta$ fields in the
intermediate effective action since, as we shall see explicitly, some of the terms in \dbiads\ result from their
thermal expectation values.

So what is the form of the effective action for $\phi$ and $\eta$? Since the dual theory is at strong coupling,
we have little hope of determining it from first principles. Nevertheless, the form of the DBI action gives us
some clues about its general structure. In the microcanonical description of the black brane solution with a
fixed energy density ${\varepsilon}$, the quantity $r_0^4/l_s^8\equiv\phi_0^4$ scales like $g_s^2\varepsilon$.
In general, the action \dbiads\ arises from couplings of $(\partial\phi)^2$ to operators made from the $\eta$
degrees of freedom -- including contributions suppressed by powers of $\phi$ -- with the $\eta$-dependent operators
thermally averaged. For example, the overall factor of $\sqrt{h(r)}$ in \dbiads\ arises from the thermal
average of a term of the form
\eqn\hoperator{
\sqrt{h(r)}\rightarrow\sqrt{1-{g_s^2{\cal O}(\eta)\over \phi^4}}+\ldots
}
where ${\cal O}(\eta)$ is an operator of dimension four whose thermal average is proportional to $\varepsilon$
(e.g. the Lagrange density itself in the ${\cal N}=4$ theory) and the
%
%
$\ldots$ represents terms which are subdominant in the black brane background. (This aspect is analogous to the
case of the DBI action itself, which also does not sum all the higher derivative terms in the theory but
contains those which dominate for particular configurations, those for which proper acceleration is small.). In
fact, in the following section, by comparing to low-temperature, weak-coupling results, we will see that ${\cal
O}$ includes
\eqn\opstart{ {\cal O}(\eta) = {1\over g_s}\,\Tr\,(\partial\eta)^2+\tilde{\cal O}}
Since this term is part of the Hamiltonian, it indeed satisfies $\langle{\cal O}\rangle_T=\varepsilon$ as
required. The remainder $\tilde{\cal O}$ represents other contributions which contribute to the energy density
in the interacting theory. This form of the action exhibits an effective ``speed limit'' on the quantity
$g_s^2{\cal O}(\eta)/\phi^4$ which is similar to the speed limit on the probe brane discussed in \KL\ and
\Dccel. Analyticity of the action bounds this operator by unity, a limiting value which upon thermal averaging
translates into $\phi > \phi_0$; this is simply the requirement that the probe remain outside the Schwarzchild
radius of the black brane created by the energy density contained in the $\eta$ excitations.

%

As organized in \hoperator, the action contains important multitrace contributions. Such terms were discussed in
\multitrace\ and are generically generated in large $N$ field theories where they contribute at leading order in
the large $N$ expansion. So although such multitrace contributions were not included in the proposals for a
``nonabelian DBI" action \nonabeliandbi, we expect them to play a role in generic backgrounds of the theory
including the thermal system. Finally, we note that the full effective action could involve an infinite series
of couplings of higher dimension operators ${\cal O}_I$ to derivatives of $\phi$ such that the summed and
averaged result agrees with \dbiads. However, at least in the ${\cal N}=4$ example, there is not an infinite set
of chiral operators which are singlets under the $SO(6)$ global symmetry and can play this role.

\subsec{Probes at Low Temperature}

In general, we cannot compute the couplings present in ${\cal O}(\eta)$ independently since the field
theory is at strong coupling. However, it will prove instructive to expand the DBI action \dbiads\ in the
regime where the probe is far from the horizon since we will be able to explain the resulting perturbation
series in $T^4/E^4$ in terms of the zero-temperature theory. Ultimately however, the regime of interest will
be the opposite one, $T \gg E$, since this regime exists also in the
$dS_d$ case.

At leading order in $r_0/r$, the dictionary between the scalar field vacuum expectation value (vev) $\phi$ and
the radial position $r$ of the probe brane is the same as in the zero temperature theory, $\phi = r/l_s^2$.  The
DBI action has both a derivative expansion and an expansion in $\lambda^2T^4/\phi^4\sim T^4/E^4$.
%
%
At zero temperature, the DBI action contains higher derivative terms which may be understood as arising from
integrating out the $\chi$ and $\eta$ modes.  In the finite temperature system, the fields in the theory,
including the $\eta$ degrees of freedom which contribute the bulk of the entropy, are excited.  As a result, the
corrections in the thermal DBI \dbiads\ arise from taking the higher derivative action including both $\phi$ and
$\eta$, and thermally averaging over the $\eta$ contributions to obtain an effective action for $\phi$.  (The
$\chi$ mode number and energy densities are highly Boltzmann suppressed in the regime we work.)


To see this in more detail, consider the leading order four-derivative interaction arising in the expansion of
the DBI action \dbiads
\eqn\fourderiv{ {\cal L}_{four-deriv}\sim N {(\partial\phi)^4\over m_\chi^4} }
which, at weak coupling, arises from integrating out the $\chi$-modes at one-loop. Here the $N$ factor reflects
the fact that we have integrated out $N=\partial S/\partial N$ $\chi$-modes, where $S\sim N^2$ is the entropy
of the system. This number $\partial S/\partial N$ of species running in the loop generating \fourderiv\ corresponds
to the difference in the entropy of the system with the brane included versus that when the brane is pulled out of the
system \DS.  Interestingly, the appearance of $\partial S/\partial N$ also appears in the analogous expansion
for the M-branes with their well-known peculiar entropies as we show in the appendix. In
this way, the probe action for a single scalar degree of freedom contains information about the entropy of the
full system.

The crucial point for our analysis of the thermal case is that the interaction \fourderiv\ implies other
interactions involving the $\eta$-modes. To see this, suppose that we are on the full Coulomb branch,
$\langle\Phi\rangle={\rm diag}(\phi_1,\ldots,\phi_N)$ so that $U(N)\rightarrow U(1)^N$. Then the generalisation
of the four-derivative terms that arises at one-loop is \eqn\dunno{ {\cal
L}_{four-deriv}\sim\sum_{i<j}{(\partial\phi_i-\partial\phi_j)^4\over (\phi_i-\phi_j)^4} } Now as we return to
our probe set-up, sending all but $N-1$ of the eigenvalues back to the origin, this coupling reduces to
\eqn\noiknow{ {\cal L}_{four-deriv}\sim N{(\partial\phi)^4\over m_\chi^4} + {(\partial \phi)^2\,\Tr (\partial
\eta)^2\over m_\chi^4} +{\Tr\,(\partial \eta)^4\over m_\chi^4} }
At this order, which is
protected by supersymmetry, the scalings \noiknow\ follow simply from the cylinder diagrams connecting the
$\phi$ and $\eta$ sectors, including the nonplanar amplitude generating the double-trace second term in
\noiknow.  In fact, these terms can be understood directly in the gravity side, without recourse to field
theory, simply by considering the couplings between energy density coming from excited $\phi$ modes on the probe
and energy density from bulk modes. As described in Section 3.1, evaluating the $\Tr(\partial\eta)^2$ term
in the thermal background correctly reproduces the thermal wavefunction renormalization seen in the
DBI action \dbiads.
%
%
%
%
We may also perform a similar trick for the $(\partial\eta)^4$ coupling. A quick evaluation of 't Hooft double
line diagrams is sufficient to see that the $N$ scaling goes as $\langle\Tr(\partial\eta)^4\rangle\sim
(1/N)\,\langle\Tr(\partial\eta)^2\rangle^2 \sim \varepsilon\,(\partial\varepsilon/\partial N)\sim N^3 T^8$ which
indeed reproduces the leading order correction to the potential in the expansion of \dbiads. This last
observation was also made by Tseytlin and Yankielowicz \tseyank.  This pattern continues to higher orders; for
example the six-derivative couplings between $\phi$ and $\eta$ generate the appropriate corrections upon
thermally averaging over the $\eta$ degrees of freedom.

\subsec{Probes at High Temperature}

In the previous sections we saw that the DBI probe action contains information about the thermal background of
the remaining $U(N-1)$ sector of the gauge theory. We would now like to study this in the low energy regime in
which the probe is close to the horizon. However, in order to do this it is necessary to sum up the higher derivative
corrections in the $\eta$ sector. To see this, note that these include a power series in the combination
$g_s^2(\del\eta)^2/\phi^4$ which, given the thermal average of \opstart, is approaching $1$ as $\phi\to\phi_0$.

Consider the action involving the $\eta$ degrees of freedom which, when thermally averaged,
reproduces the probe DBI action \dbiads.
If we now expand the thermally averaged action about $\phi=\phi_0$, we obtain an effective action for the probe
appropriate to small $E/T$.  In this limit,
\eqn\hlim{h=1-{\langle {\cal O}(\eta)\rangle g_s^2\over \phi_0^4} \to {{4(\phi-\phi_0)}\over\phi_0}}
It is this regime near $\phi=\phi_0$ which generalizes to the de Sitter case.  The leading behavior of the probe
action in this regime arises in the field theory from a nontrivial sum of higher derivative terms generated near
the origin of the approximate moduli space.

In this section, we have focused on the specific case of the ${\cal N}=4$ super Yang-Mills theory, for which the
degrees of freedom $(\phi,\chi,\eta)$ make up a $U(N)$ gauge theory.  In general, for gravity duals of $4d$
conformal field theories the geometry is as in \adsfive\ and hence the square root term in the probe DBI action
takes the form in \dbiads. The couplings encoded in \noiknow, leading to \dbiads, are also still
present, but their microscopic origin will differ in different cases.  Similarly, in other dimensions (even with
maximal supersymmetry) the same interpretation arises though the couplings cannot be calculated perturbatively
in any limit.  The couplings \noiknow\ could for example be analyzed directly on the gravity side of the
correspondence, by considering the Coulomb branch configuration \dunno, since they represent couplings between
$\phi$ and $\eta$ energy densities mediated by bulk fields.  We will not pursue this independent calculation
here.

\subsec{Probe Dynamics}

Before returning to the de Sitter case, let us here note some amusing features of the probe dynamics in this
thermal theory.

Firstly, there is a stronger speed limit generated by the thermal background, generalizing the case studied in
\refs{\KL,\Dccel,\DBIsky}. The motion of $\phi$ determined from the equations of motion of \dbiads\ has the
property that $\phi$ takes forever to reach $\phi_0$.  The speed limit on $\phi$ motion here is
\eqn\thermalspeedlimit{|\dot\phi|\le {\phi^2\over\sqrt{\lambda}}\left(1-{\phi_0^4\over\phi^4}\right)}
leading to a late-time solution
\eqn\latesol{\phi-\phi_0 \sim ce^{-4\phi_0 t/\sqrt{\lambda}}}

Again, this behavior is due to the strong back reaction on $\phi$'s motion from the higher derivative couplings
to the thermally $\eta$s via the $\chi$ modes.  Thus the somewhat counterintuitive fact that the probe takes
forever to reach the horizon in GR is holographically dual to a pure quantum field theory effect:  the strong
back reaction on rolling scalar fields as they approach points with light states on their approximate moduli
space.  Again, this generalizes the effects studied in \Dccel\ and \trapping\ to the thermal case where both a
gas of particles and higher derivative corrections play crucial roles.

Secondly, let us remark on a related phenomenon that is predicted by the AdS/CFT correspondence. Consider the
zero-temperature field theory with a few eigenvalues out on its Coulomb branch, as described on the gravity side
by few D3-brane probes.  As a localized quantum mechanical system, the probe branes constitute an Unruh
detector.  So if the probe branes are forced onto a trajectory with nontrivial proper acceleration $a_p$ on the
gravity side of the correspondence, they will register the corresponding temperature $T=a_p/2\pi$.  This makes a
prediction that the corresponding motion on the field theory Coulomb branch should lead to thermal Greens
functions to a good approximation at low energies in the sector of the theory corresponding to the set of
eigenvalues out on the Coulomb branch.

\newsec{dS Redux}

In this section, we return to the study the dynamics of D-brane probes on $dS_{d-1}$ slices of $(A)dS$ space,
using the lessons learned in \S3.  This will enable us to flesh out some further aspects of the correspondence
described in \S1\ and \S2.  In particular, the dynamics of the probes representing a field theory scalar field
out on its Coulomb branch encodes aspects of the relevant higher derivative couplings in the dual field theory
at low energies, as we summarize in \S4.1.  As we review in \S4.2, going out fully on the Coulomb branch
provides a way to determine the microphysical content of the dual CFT from brane probes.  Brane probes in other
configurations, describing domain walls in the dual $d-1$ dimensional theory, are discussed in \S4.3.  Finally,
in \S4.4\ we discuss the prospects for pursuing our duality further to higher codimension, ultimately decoupling
the description from dynamical gravity.

\subsec{The Coulomb branch of the dual CFT and higher derivative couplings}

The Coulomb branch dynamics of our field theory can be explored through the use of a brane probes of the
metric \dSslices\ at position $\rho$, governed by the DBI action
\eqn\DBIdSslices{S_{DBI}=-\tau_B\int d^{d-1}y\sqrt{\hat g}\left({\rho\over R}\right)^{d-1} \sqrt{1+{R^4\over
\rho^2}{{\hat g^{\alpha\beta}\partial_\alpha \rho\partial_\beta \rho}\over{R^2\pm \rho^2}}}+S_{WZ}}
where $\tau_B$ is the brane tension and the $\pm$ refers to $AdS$ and $dS$ respectively.

The Wess-Zumino term $S_{WZ}$ depends on the flux quantum numbers and corresponding probe brane charges.  For
example, in the case of $AdS_5$ with $dS_4$ slices, the WZ term is
%
%
%
\eqn\WZAdS{{\cal S}_{WZ,AdS}=c_{AdS}\left( {\rho\over R}\sqrt{1+{\rho^2\over R^2}}(-3+2{\rho^2\over R^2})+3{\rm
ArcSinh}{\rho\over R} \right)}
where the constant prefactor $c_{AdS}$ depends on the flux and brane charge content of the background and the
probe.   Similarly, for $dS_5$ with $dS_4$ slices we obtain
\eqn\WZdS{{\cal S}_{WZ,dS}=c_{dS}\left( -{\rho\over R}\sqrt{1-{\rho^2\over R^2}}(3+2{\rho^2\over
R^2})+3\rm{ArcSin}{\rho\over R} \right)}

The interpretation of the terms in the DBI action is
similar to the case of thermal field theory on Minkowski space described in \S3 for temperature
$T=1/R$ (that is setting $r_0=R$ in the formulas of \S3). Again the DBI action arises from the summation of
higher derivative couplings (suppressed by the $\chi$ mass scale $\phi$), evaluated in the background of
interest (in this case $dS_{d-1}$ with its curvature radius and inverse temperature $R$).

However the two are not exactly the same.  For instance, in the case of $AdS_d$, expanding \DBIdSslices\ about
large $\rho$ one finds corrections of order $T^2/E^2$, whereas in the flat space thermal field theory \dbiads,
the corrections arose in a power series in $T^4/E^4$ at large $r$.  Also, in the case of conformal field theory
on $dS_{d-1}$, the conformal coupling proportional to ${\cal R}\phi^2$ is nontrivial in the background
\swconformal, whereas its coefficient ${\cal R}$ vanishes in the flat space case.  Moreover, there is a whole
series of corrections in ${\cal R}/\phi^2$ (multiplying couplings) arising in a space of general curvature
\swconformal.  In fact there is a relation between these two statements; the curvature couplings generate extra
corrections scaling like $\lambda {\cal R}l_s^4/\rho^2 \sim T^2/E^2$.\foot{To avoid confusion let us note that
$\rho$ variable is different from the $\phi l_s^2$ variable of \swconformal\ away from the boundary.}
%
%
%
%
Again, these couplings carry information about the $N=\partial S/\partial N$ degrees of freedom in the $\chi$
modes coupling the probe with the rest of the system.  Another distinction is the logarithmic contributions in
the Wess-Zumino term \WZAdS, which would be interesting to interpret in the field theory.

Away from the high energy limit, it is useful to change variables from $\rho$ to $r_T$, where $r_T=\phi_T l_s^2$
is similar to the $r=\phi l_s^2$ variable of the thermal flat space field theory case of \S3:
\eqn\newvar{\rho=2\sqrt{R(r_T-R)}}
In terms of this variable, the metric near the center of the $dS_{d-1}$ probe brane is approximately equal to
\adsfive, with $r\sim r_T$ and $h=1-R^4/r_T^4$.  As the probe approaches the horizon at $\rho=0$, $r_T$
approaches $R$.  In this limit, consider the quantities
\eqn\hdone{
{g_s^2{\cal O}(\eta) \over \phi_T^4}\ \ \ \ \ \ {\rm and}\ \ \ \ \ \ \
{\lambda {\cal R}\over\phi_T^2}}
which appear in the higher derivative expansion of the action (the second one arising from the curvature
coupling in the $dS_{d-1}$ slicing as just described).   Since they are approaching 1, similarly to the case of
\S3, the behavior near the horizon involves a nontrivial sum of higher derivative terms correcting the moduli
space approximation.

Altogether, taking into account these modifications, the probe action \DBIdSslices\ carries information about
the nontrivial higher derivative couplings to the $S$ microscopic $\eta$ degrees of freedom in the same way as
described in \S3\ (though as discussed at the end of \S3.3, the microphysical origin of these couplings will
vary from case to case). Again, the behavior of the probes in the $E\ll T$ regime is a pure quantum field
theory effect.

\subsec{The Content of the CFT on the Coulomb Branch}

Although we have elucidated the structure of codimension 1 holography in the de Sitter static patch, it is of
great interest to determine the microphysical content of the dual thermal field theory in specific examples,
such as \refs{\mss,\kklt}. In the AdS/CFT correspondence, one has a general dictionary
\refs{\anothercitation,\adsdict}\ which is independent of the specific content of the dual field theory and
which is indeed realized by many different microphysical examples such as the ${\cal N}=4$ SYM theory. We
expect a similar situation here:  the general structure of the correspondence we developed here and in
\refs{\DS,\karch}\ should apply in many different microphysical realizations of dS space.

Let us briefly review here how this content might be determined, again using the branes available on the gravity
side of the correspondence to deduce features of the field theory side.  Namely, in de Sitter flux models such
as \refs{\mss,\kklt}, one can go fully onto the Coulomb branch by trading all of the flux in the IR end of the
geometry for branes \refs{\DS,\junctions}.  In the examples of \kklt, one finds on the Coulomb branch of the
system a product of unitary gauge groups with multifundamental matter degrees of freedom arising from string
junctions \junctions, whose entropy matches that predicted in \counting\ assuming a roughly uniform distribution
of cosmological constants. These examples contain dynamical wrapped NS5-branes as well as D5 and D3-branes,
which in principle determine the couplings among the multifundamental matter fields in the system.\foot{We thank
O. de Wolfe for ongoing discussions on this system.} In the cases \mss, one can again trade the flux for
branes out on the Coulomb branch; in this case one can avoid $NS$ branes and have only D-branes, with the
dilaton tadpole from the supercriticality providing the independent force analogous to the NS flux in \kklt. It
will be interesting to pursue these specific examples further, to see how the microphysics in each case fits
into the general dictionary we outlined here.

Note that out on the Coulomb branch, the induced contribution to the $d-1$ dimensional Newton constant is
modified from its value \rennewtII\rennewtI.  On both sides of the duality, the renormalized contribution is
reduced out on the Coulomb branch.  In the $d-1$ description, this happens because low energy field theory modes
are massed-up on the Coulomb branch, and hence contribute less to the total renormalization of $G_N$. Similarly
on the gravity side, trading the flux near the horizon for branes leads to a smaller geometrical contribution to
$G_N$ because the region near the horizon is replaced by one with smaller $d$-dimensional cosmological constant
and hence less volume.

\subsec{Domain walls and vacuum bubbles}

In addition to the probe branes living on the $dS_{d-1}$ slices at fixed $\rho$, which manifest the Coulomb
branch of the field theory, we can consider other configurations of codimension one brane probes.  Another
simple case is a probe arranged as a shell surrounding the static observer.  This
probe spatially traces out a
sphere at a (time dependent) radius $r_s(t)$ in the static coordinate system
\eqn\dSstatic{ds^2=-(1-r_s^2/R^2)dt^2+(1-r_s^2/R^2)^{-1}dr_s^2+r_s^2d\Omega^2}
\ifig\vacbubble{A domain wall shell in the static patch.  Note that the bubble wall intersects the $dS_{d-1}$
slices on which the field theory lives also as a vacuum bubble in the field theory.}
{\epsfxsize2.5in\epsfbox{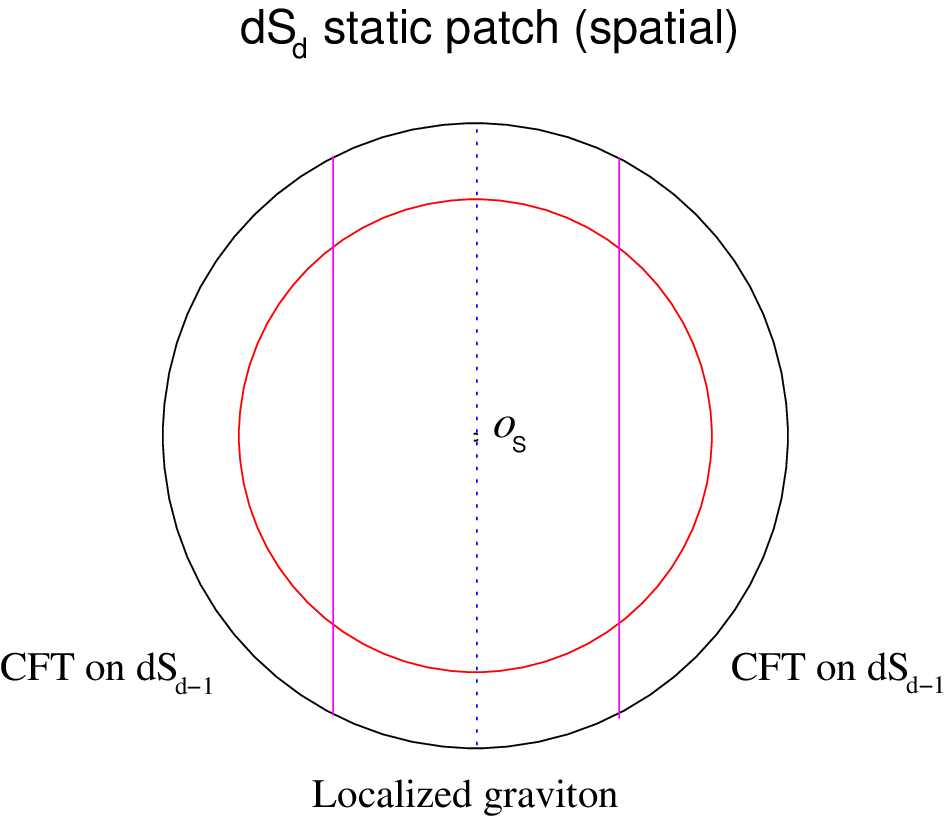}}
As in the thermal field theory case discussed in \S3, this probe takes forever to fall into the $dS_d$ horizon
at $r_s=R$ in the general relativity approximation.  Again this follows from a combination of summed higher
derivative and thermal effects in the probe DBI action
\eqn\DBIraddS{{\cal S}_{DBI}=-\tau_B\int r_s^2d^2\Omega dt \sqrt{h}\sqrt{1-\dot r_s^2/h^2}}
where $\tau_B$ is the tension of the brane probe.

Now let us consider the interpretation of these branes in the dual.  By superimposing this brane on the
$dS_{d-1}$ slicing defining our dual depicted in figure 1, we see that the branes \DBIraddS\ constitute domain
walls in the dual $d-1$ dimensional theory (see figure 3).  In this way, the vacuum structure of the discretuum
of gravity solutions should be dual to a discretuum of vacua of the dual.  It would be interesting to pursue
this connection further, especially given the arguments for decays of existing de Sitter constructions in string
theory \refs{\kklt,\mss}.  Note also that the duality between the pure dS and the cutoff CFT is unaffected by
such decays for exponentially long timescales during which the simpler setup we have focused on in this paper is
a good approximation.  The domain wall configurations here extend to the UV cutoff region of the geometry, and
hence do not involve purely physics of the CFT of the simpler approximation.

\subsec{Higher Codimension and non-gravitational holography}

Combining all the results here with \refs{\DS,\karch}, we may obtain further holographic dualities by iterating
the dimensional reduction procedure \inprogress.  Namely, the localized $d-1$ dimensional graviton of \karch\
suggests that the system should further be holographically dual to a $d-2$ dimensional theory and so on, leading
to CFT on $dS_2$ (as studied in \martinec) and ultimately to quantum mechanics.  The content of this quantum
mechanical theory on its approximate moduli space can be determined from the degrees of freedom living on the
corresponding intersecting D-brane probes \refs{\DS,\junctions}.

It will be interesting to see if this quantum mechanical theory arising in this way from explicit models has any
relation to the interesting approaches to de Sitter holography suggested in \refs{\otherdS,\tom}. The proposals
\otherdS\ aim for a complete description of global asymptotically $dS_d$ space, and some of them define
observables (or ``calculables") along one or both of the spacelike boundaries of $dS_d$.\foot{though these may
not exist in the same form after all decay processes are taken into account \refs{\kklt,\lennyanth}.}

In our correspondence, we confine our description to a causal region, with the holographic dual defined on
timelike slices amenable to explicit analysis of the Coulomb branch via branes traded for flux in stringy de
Sitter constructions \refs{\mss,\kklt}. One might wonder about the precision of a holographic duality defined in
a geodesically incomplete patch of a spacetime. This situation arises also in the Poincare patch version of
$AdS/CFT$, whose field theory side is conformal field theory on Minkowski space. It is this version of $AdS/CFT$
one that is more closely analogous to the causal patch holography for de Sitter space.  In both cases, although
the horizon is only an approximate general relativistic concept, the system is defined quantum mechanically by
the dual field theory.

In the $AdS$ case, there are several versions of the correspondence pertaining to different patches of the space
(the Poincare patch being dual to CFT on Minkowski space, the global $AdS$ being dual to CFT on $S^3\times \IR$,
and the plane wave limit being dual to a quantum mechanical theory \shahin).  It is entirely possible that de
Sitter holography works the same way, with different mutually consistent versions.  As just mentioned, the
Poincare patch version of $AdS/CFT$ is probably the closest analogue to the static patch of $dS$, as the gravity
side is geodesically incomplete and one must take into account the processes that enter the horizon in the far
past and exit in the far future.  It is clear that even if there is a global version of de Sitter holography,
there must be a self contained description of what a given observer can access, which the present proposal
provides.

\medskip

\noindent{\bf Acknowledgements}

We would like to thank O. Aharony, B. Frievogel, O. de Wolfe, S. Kachru, A. Lawrence, N. Kaloper, A. Maloney, R.
Myers, S. Shenker, A. Starinets, M. Strassler, A. Strominger, R. Sundrum, L. Susskind, and S. Trivedi for very
useful discussions. We would like to thank the organizers and participants of the Banff 2003 and 2004, CERN
2004, Johns Hopkins 2004, and Texas A. \& M. 2004 workshops where various stages of this work were presented.
M.~A. is supported in part by Iranian TWAS chapter based at ISMO, and thanks the ICTP for hospitality during the
completion of this work. AK is supported in part by DOE contract DE-FG02-96ER40956. E.~S. is supported in part
by the DOE under contract DE-AC03-76SF00515 and by the NSF under contract 9870115. D.T. is supported by a
Pappalardo Fellowship and is grateful the Pappalardo family for their generosity. D.T. would further like to
thank SLAC, the Department of Physics at Stanford University, and the Research Institute for Mathematical
Science at Kyoto University for their kind hospitality while this work was undertaken.

\appendix{A}{M-branes}

In Section 3, we interpreted the expansion of the D3-brane probe action in terms of an underlying thermal field
theory. From the discussion there, we found the the low-temperature expansion of the black D3-brane probe
can be written in the suggestive form,
\eqn\newexpan{ {\cal L}_{DBI}\sim {1\over g_s}{\partial {\varepsilon}\over
\partial N}\left(1+g_s^2{{\varepsilon}\over m_\chi^4}+\ldots\right) +
{1\over g_s}\left(1+g_s^2 {{\varepsilon}\over m_\chi^4}+\ldots\right)
(\partial \phi)^2+{\partial S \over\partial N}{(\partial\phi)^4\over m_\chi^4}+\ldots }
In fact, the same interpretation can be shown to work for other, non-conformal, D-branes as well as
orbifold field theories. In this appendix we show how a similar expansion appears for the black
M-branes where, from the perspective of the dual, strongly coupled, CFT, the terms in \newexpan\ arise
from a combination of quantum and thermal effects. In this manner, the probe
action knows about the number of species $S$ present in the dual field theory in much the same way that
the Z-boson knows about the number of light neutrino species.

\subsec{M2-Brane}

The near horizon limit of $N$ non-extremal M2-branes yields the 7-sphere with $N$ units of flux times the
$AdS_4$-Schwarzchild metric, where the latter is given by,
\eqn\mtwometric{ ds^2={r^4\over
R^4}(-h(r)dt^2+d\vec{x}^2)+ {R^2\over r^2} h(r)^{-1}dr^2 }
where the non-extremality function is given by $h(r)=1-{r_0^6/r^6}$. The dictionary between the field theory and
supergravity variables includes the entries $(MR)^9\sim N^{3/2}$ and $r_0^2=TR^3$ where $M$ is the
11-dimensional Planck scale. Using supergravity, one can show that the free energy of this solution is given by
${\cal F}\sim N^{3/2}T^3$, which exhibits the well-known and ill-understood $N^{3/2}$ scaling of the entropy. We
will now see how, with some interpretation of the probe action, one may rederive this result for the scaling of
the entropy at zero temperature.

The action for a probe M2-brane in the background \mtwometric\ is given by
\eqn\probemtwo{ S_{M2}=-M^3\int d^3x\ {r^6\over R^6}\,\left(h^{1/2}\,\sqrt{1-{R^6\dot{r}^2\over r^6 h^2}
+{R^6(\nabla r)^2\over r^6 h}}-1\right) }
from which we can extract the four-derivative terms at zero temperature. Expressing these in terms of the scalar
field $\phi=rM^{3/2}$, which has engineering dimension $[\phi]=1/2$, we have,
\eqn\fourmtwo{ {\cal L}_{4-deriv}
\sim M^3R^6\frac{(\partial{r})^4}{r^6}\sim N\frac{(\partial{\phi})^4}{\phi^6} }
which comes with a coefficient of $N$, as in the case of the D3-brane, giving no hint of the $N^{3/2}$ degrees of
freedom living on the M2-brane
worldvolume. However, suppose that this term can be thought of as arising in the M2-brane theory by integrating
out ``$\chi$-modes as in the D3-brane.  In this more general setting, we do not have a perturbative description
of the field theory in any regime (other than the large radius low energy description on the gravity side of the
AdS/CFT correspondence), so we do not have a simple diagrammatic interpretation of the counting of ``$\chi$
modes".  However rather generally we expect their number to be simply the difference $\partial S/\partial N$ between
the number of degrees of freedom of the system with one probe removed from the stack of $M2$-branes and the
number of degrees of freedom in the system with the probe included in the stack \DS.  Indeed we will find this
scaling appearing in the DBI action for the M-branes in the same way it appeared for the D3-branes.  Here the
$\chi$ modes, stretching between M2-branes, are best thought of as bion-like spikes in the M2-brane
worldvolume. Since the only scale in town in the Planck mass $M$, the tension of these objects must scale as
$M^2$ in flat space. Then, in a pure $AdS_4$ background (or for large vevs in $AdS_4$-Schwarzchild),
the mass of these $\chi$ modes is \eqn\mchimass{ m_\chi = M^2 \int\ dr\ \sqrt{g_{00}}\sqrt{g_{rr}} = M^2\int\ dr
\ {r\over R} = M^2 {r^2\over R} } We may now rewrite the four-derivative interaction \fourmtwo\ in terms of
$m_\chi$
\eqn\again{ {\cal L}_{4-deriv}\sim (RM)^3 {(\partial{\phi})^4\over m_\chi^3} =N^{1/2}{(\partial{\phi})^4\over
m_\chi^3} } to see that the coefficient indeed captures the conjectured number of $\chi$-modes that we integrated
out: $\partial S/\partial N=N^{1/2}$.

Let's now turn to finite temperature effects and repeat the analysis that we performed in Section 2 for the
D3-branes. The leading order expansion of the DBI action includes the couplings
\eqn\mtwoexpan{ S_{M2}\sim \int d^3x\ N^{1/2}T^3\left(1+{N^{3/2}T^3\over \phi^6}+\ldots\right)
+\left(1+{N^{3/2}T^3\over \phi^6}+\ldots\right)(\partial\phi)^2+\ldots }
which matches the general form of the expansion given in \newexpan. As for the D3-branes, we would like to
understand how these terms arise from the couplings of $\phi$ to the bulk of the system (whose degrees of
freedom we will again refer to as $\eta$ modes) via $\chi$s.  However, we do not have a perturbative field
theory description in which we can calculate the contributions directly.  In the D3-brane case, we could also
determine this four derivative coupling from the closed string channel, whose low energy contribution is
supergraviton exchange.  This coupling is simply controlled by Newton's constant (in $10d$ for the D3-brane
case) and has no extra powers of $N$.  This generalizes to the current situation--from the gravity point of
view, the coupling \fourmtwo\ generalizes to contributions
$(\partial\eta)^2(\partial\phi)^2/\phi^6+(\partial\eta)^4/\phi^4$. In terms of $m_\chi$, this translates into
the couplings
\eqn\againagain{ {\cal L}_{4-deriv}\sim \sqrt{N}\,{(\partial{\phi})^4\over m_\chi^3}
+{1\over\sqrt{N}}{\Tr(\partial\eta)^2\,(\partial\phi)^2\over m_\chi^3}+{1\over\sqrt{N}}
{\Tr(\partial\eta)^4\over m_\chi^3} }
where, by ``Trace'', we mean a suitable sum over all degrees of freedom of the M2-brane worldvolume. In
particular, the Hamiltonian of the M2-brane theory -- whatever it is! -- includes the term $\Tr
(\partial\eta)^2$, so that when we evaluate this in the background thermal bath we have
$\langle\Tr(\partial\eta)^2\rangle \sim F\sim N^{3/2} T^3$. In this way, the second term in \againagain\ can be
shown to reproduce the thermal wavefunction renormalisation arising from the expansion of the DBI action
\mtwoexpan. Following the D3-brane example, we expect that the correction to the free-energy in \mtwoexpan\
arises from calculating the $\Tr(\partial\eta)^4$ coupling in the thermal background. Our lack of understanding
of the M2-brane worldvolume prohibits a first principles computation of this expectation value. Here we note
that we get the right functional form provided the answer scales in the manner:
$\langle\Tr(\partial\eta)^4\rangle\sim (1/N)\langle(\Tr\partial \eta)^2\rangle^2$. We may take this as a clue as
to the structure of the non-abelian gauge theory on the M2-brane.

\subsec{M5-Brane}

The same pattern holds for the M5-brane, using exactly the same logic as in the $M2$ case but replacing the warp
factor and thermodynamic quantities with the powers of $r,R$ and $N$ appropriate to this case.  Again, to
find the quantity $\partial S/\partial N$ in front of the four-derivative
correction, one must normalize by dividing by a suitable $m_\chi$. The relevant $\chi$-modes are M2-branes
stretched between M5-branes. If these M2-branes are curled on a cylinder of radius $1/M$ (again: there is
no other scale) then the tension of the ``$\chi$-string'' is $M^2$ and this leads to the correct scaling
$\partial S/\partial N\sim N^2$.

\listrefs

\end